\documentclass[amsmath,notitlepage,aps,prd,10pt,onecolumn]{revtex4-1}

\usepackage{amssymb}
\usepackage{slashed}
\usepackage{txfonts}
\usepackage{graphicx}
\usepackage{xcolor}
\usepackage{titlesec}
\usepackage{subfig}
\usepackage{multirow}
\usepackage{hhline}
\usepackage{lipsum}
\usepackage{amsmath}
\usepackage{hyperref}
\usepackage{tabulary}

\begin{document}
\title{Production of $J/\psi + \chi_c$ and $J/\psi + J/\psi$ with real gluon emission at LHC}
\author{A.K. Likhoded}
\email{Anatolii.Likhoded@ihep.ru}
\affiliation{Moscow Institute of Physics and Technology, Dolgoprodny, Russia}
\affiliation{Institute for High Energy Physics NRC ``Kurchatov Institute'', 142281, Protvino, Russia}
\author{A.V. Luchinsky}
\email{Alexey.Luchinsky@ihep.ru}
\author{S.V. Poslavsky}
\email{stvlpos@mail.ru}
\affiliation{Institute for High Energy Physics NRC ``Kurchatov Institute'', 142281, Protvino, Russia}

\begin{abstract}
In the present work we study production of $J/\psi + \chi_c$ and $J/\psi + J/\psi$ in LHC. The first process is forbidden at the leading order in gluon fusion due to the $C$-parity conservation, and the first non-vanishing contribution is given by the process with additional emission of real gluon. Considering the direct production of $J/\psi + J/\psi$, where the leading order is allowed, we have found that the contribution from the higher order process with real gluon emission is comparable and even more significant than the leading order. Moreover, account of this higher order effect dramatically changes kinematical distributions. Through the present paper we study in details different channels of paired $J/\psi + J/\psi$ production: direct $J/\psi + J/\psi$ production, feed-down from $J/\psi + \chi_c$ channel, double parton scattering. We also try to find kinematical distributions that are most suitable to separate these different channels.
\end{abstract}

\maketitle

\newcommand{\Q}{\mathcal{Q}}

\newcommand\nb{\mbox{nb}}
\newcommand\pb{\mbox{pb}}
\newcommand\GeV{\mbox{GeV}}
\newcommand\TeV{\mbox{TeV}}

\newcommand{\QQQ}{\fbox{???}}
\newcommand{\OS}{\left<O_S\right>}\newcommand{\OP}{\left<O_P\right>}
\newcommand{\EmptyBox}{\centering\fbox{\begin{minipage}{3in}\hfill\vspace{3in}\end{minipage}}}

\section{Introduction}
Heavy quarkonia production in high energy hadronic interactions is a perfect instrument for rigorous studies of QCD nature. Due to the large mass of heavy quark, production of quark-antiquark pair is ruled by the perturbative QCD, while its hadronization can also be described theoretically due to the nonrelativistic nature of these bound states. Moreover, small width of heavy quarkonia provides clean experimental signature, so there is a big experimental statistics available. 

Multiple production of heavy quarkonia, like paired production of $J/\psi$, attracts additional interest due to the possibility to study parton shower structure in details, since double parton scattering (DPS) may play a tangible role in these processes \cite{Kom:2011bd, Novoselov:2011ff, Baranov:2012re}. In this connection, paired $J/\psi$ production was studied theoretically by a number of authors \cite{Li:2009ug, Ko:2010xy, Berezhnoy:2011xy, Berezhnoy:2012xq, Li:2013csa, Lansberg:2013qka, Lansberg:2014swa, Sun:2014gca}, and, in general, one can conclude that single parton scattering (SPS) does not underestimate nor contradict experimental cross sections measured by LHCb \cite{Aaij:2011yc}, CMS \cite{Khachatryan:2014iia} and D0 \cite{Abazov:2014qba} collaborations. It is worth to mention, however, that for the multiple production involving open charm contradictions of SPS predictions with the experimental data are considerable \cite{Berezhnoy:2012xq, Maciula:2013kd}.

In the situation when we cannot certainly discriminate between different channels in paired heavy quarkonia production, there is a promising possibility to shed light on this problem by considering final states that are forbidden or suppressed in the leading order SPS, but can easily be produced in DPS. Recently such processes as $\Upsilon + J/\psi$ and $\Upsilon + c\bar c$ were considered in a series of works \cite{Likhoded:2015zna, Berezhnoy:2015jga, Likhoded:2015fdr}, where it was shown that NLO effects and feed-down from $P$-wave states (i.e. $\chi_b + \chi_c$ and $\chi_b + c\bar c$) contribute significantly to the SPS picture. Moreover, it was shown that the feed-down from $\chi_b + c\bar c$ to associated $\Upsilon$ production obtained within SPS approach has very similar shapes of the differential distributions to that were observed by the LHCb \cite{Aaij:2015wpa} except for the azimuthal asymmetry, which is the most distinguishing feature of the DPS.

In the charm sector forbidden within LO SPS final state  is $J/\psi + \chi_c$, since in gluon fusion $gg \to 1^{--} + J^{++}$ is forbidden due to $C$-parity. The first non-vanishing contribution at the  parton level is thus given by the process with additional radiation of real gluon which is $O(\alpha_S^5)$. Such additional emission was first considered for direct $J/\psi + J/\psi$ production in \cite{Lansberg:2013qka, Lansberg:2014swa, Sun:2014gca}, and it was shown that even in this case which is allowed at LO, the $O(\alpha_S^5)$ corrections are very important. As it will be shown in the rest of the paper, at $\sqrt{s} = 13\, \TeV$ these corrections are even more significant than the LO. It will be also shown that by switching between different final states $J/\psi + J/\psi$ or $J/\psi + \chi_c$ we thereby switch between different underlying mechanisms of the hard reaction. Finally, it is worth to point out that considering additional radiation of real gluon solves another serious problem of LO SPS approach: at the LO the hard reaction $gg\to J/\psi + J/\psi$ has strong back-to-back kinematics, so using standard collinear parton distributions (PDFs) it is not possible to describe azimuthal or $p_T$ asymmetries or non-vanishing $p_T$ of a whole pair which are observed experimentally. In the case of $J/\psi + J/\psi$ this could be overcomed by using unintegrated PDFs in the model of $k_T$-factorization \cite{Baranov:2015cle}, while $J/\psi + \chi_c$ is still forbidden at LO.

The rest of the paper is organized as follows. In the next section we discuss main features of the parton processes $gg \to J/\psi + \chi_c + g$ and $gg \to J/\psi + J/\psi + g$. In Section III we give theoretical predictions for hadronic cross sections. Brief results are given in the last section.

\section{Parton level}
It is well known that in hadronic interactions at high energies the main contribution at the parton level is given by gluon-gluon fusion. Production of a pair of two heavy quarkonia $\Q_1$ and $\Q_2$ at LO is described by the process $gg \to \Q_1 + \Q_2$. Two gluons in white configuration have positive $C$-parity, which restricts possible quantum numbers of the final state and e.g. production of $J/\psi + \chi_c$ is forbidden. For this configuration it is necessary to go beyond LO and consider partonic reaction with additional emission of a gluon:
$$
gg \to \Q_1 + \Q_2 + g,
$$
which is of order $O(\alpha_S^5)$. Let us note, that in the case of $J/\psi + J/\psi$ the full $O(\alpha_S^5)$ NLO corrections include one loop contributions which are not considered in this paper since they again have only back-to-back kinematics. Following \cite{Lansberg:2013qka} we shall label $O(\alpha_S^5)$ contributions due to the real emission as NLO*.

\begin{figure}[h]
\includegraphics[width=\textwidth]{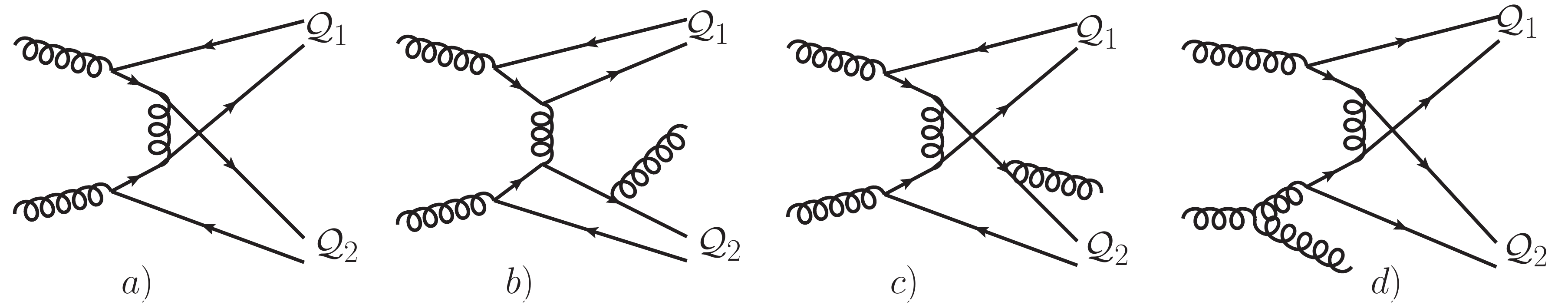}
\caption{Typical Feynman diagrams for the paired heavy quarkonia production at parton level: $a)$ diagram for the LO process (only for $J/\psi + J/\psi$), $b)-d)$ diagrams for the NLO* process (both $J/\psi + \chi_c$ and $J/\psi + J/\psi$ are possible).
\label{fig_diagrams}}
\end{figure}

Typical Feynman diagrams of the partonic reaction at LO and NLO* are shown in Fig.~\ref{fig_diagrams}. In total there are 438 Feynman diagrams (which we have generated using \verb!FeynArts! \cite{Hahn:2000kx}) for the NLO* process assuming that each final quark-antiquark pair is formed in the colour singlet combination. Here we should note, that in the following we use the leading order colour singlet term in the NRQCD velocity expansion \cite{Bodwin:1994jh} for the description of $J/\psi$ and $\chi_c$ hadronization. In the case of $J/\psi$ this is the leading term of the expansion, while for the $\chi_c$ the leading colour octet term is suppressed \cite{Likhoded:2012hw, Likhoded:2014kfa} and only singlet contribution is relevant. Within this approximation, the quark-antiquark projectors have the following form:
\begin{eqnarray}
u \left(\frac{p_1}{2}\right) \bar{v} \left(\frac{p_1}{2}\right) \quad &\equiv& \quad \Pi_S \quad =  \quad  \frac{1}{2\sqrt{6}}\, \left(\hat{p}_1 - 2 m_c\right) \hat{\epsilon}_\psi,\\
u \left(\frac{p_2}{2} + q\right) \bar{v} \left(\frac{p_2}{2} - q\right) \quad &\equiv&  \quad \Pi_P \quad = \quad  \frac{1}{8\sqrt{6} \, m_c^2} \left(\frac{\hat{p}_2}{2}-\hat{q}-m_c\right)\hat{\epsilon}_S\left(\hat{p}_2 + 2m_c\right) \left(\frac{\hat{p}_2}{2}+\hat{q}+m_c\right),
\end{eqnarray}
where the first one is a projector for $J/\psi$ and the second one for the $\chi_c$. Here $p_1$ and $\epsilon_\psi$ are momentum and polarization vector of $J/\psi$, $p_2$, $\epsilon_S$ and $q$ are momentum, spin polarization vector and relative momentum of quarks of $\chi_c$. Following this definitions, the matrix elements for double charmonia production can be written as (for the details on the projector operators see \cite{Braaten:2002fi}):
\begin{eqnarray}
\mathcal M(J/\psi + J/\psi) \quad &=& \quad \left[\sqrt{\frac{1}{4\pi m_c}} R_S(0) \right]^2 \mathcal A,\\
\mathcal M(J/\psi + \,\chi_c) \quad  &=& \quad \sqrt{\frac{1}{4\pi m_c}} R_S(0) \sqrt{\frac{3}{4\pi m_c}} R_P'(0) \left( \frac{\partial}{\partial q_\mu} \mathcal A(q) \right)_{q=0} \epsilon^\mu_L,
\end{eqnarray}
where $R_S(0)$ is $J/\psi$ radial wave function at the origin and $R'_P(0)$ is derivative of $\chi_c$ wave function, $\epsilon_L$ is angular polarization vector of $\chi_c$. The values of $R_S(0)$ and $R'_P(0)$ can be determined from the known widths of $J/\psi \to e^+e^-$ and $\chi_{c0} \to \gamma\gamma$ decays:
\begin{eqnarray*}
\Gamma(J/\psi \to e^+e^-) &=&  \frac{e_Q^2 \alpha^2}{m_c^2} R_S(0)^2, \\
\Gamma(\chi_{c0} \to \gamma\phantom{^+}\gamma\phantom{^+}) &=&  \frac{27 e_Q^4 \alpha^2}{m_c^4} R_P'(0)^2 .
\end{eqnarray*}
Using PDG \cite{Agashe:2014kda} values for these widths and $m_c = 1.5\, \GeV$ we get $R_S(0)^2 = 0.51\, \GeV^3$ and $R'_P(0)^2 = 0.042\, \GeV^5$. The latter value is too small compared to the value obtained from the fit of the single $\chi_c$ production in LHC \cite{Likhoded:2014kfa} which is $\approx 0.3\, \GeV^5$. In the rest of the paper we will use this value for the $R'_P(0)^2$.

We have calculated all matrix elements analytically using \verb!Redberry! computer algebra system \cite{Bolotin:2013qgr}, while squaring of the amplitude was performed numerically, using explicit Clebsh-Gordon coefficients for composing $\epsilon_S$ and $\epsilon_L$ into a state with particular total spin $J$.

Let us now look more closely on the properties of Feynman diagrams and first let's focus on $gg \to J/\psi + \chi_c + g$ process. In general matrix element with three gluon currents has terms  proportional to either symmetric $d_{ABC}$ or antisymmetric $f_{ABC}$ SU(3) structure constants. We have found that in this process all terms proportional to $f_{ABC}$ cancel each other, which can be explained by assigning negative charge parity to this structure. All diagrams of type \ref{fig_diagrams}b with final state radiation (FSR) from $\chi_c$ line cancel each other giving in total zero since $J/\psi(1^{--})$ can't be formed from two gluons. Additionally, all 144 diagrams with initial state radiation (ISR) (Fig.~\ref{fig_diagrams}d) also cancel each other due to the $C$-parity ($C$-odd final state can't be formed from two gluons). Next, it was explicitly checked that all FSR diagrams with gluon emitted from $J/\psi$ are infrared and collinear stable, while diagrams with gluon emitted from $\chi_c$ (Fig.~\ref{fig_diagrams}c) have collinear singularity (there are in total 53 such diagrams). It is well known, that this singularity can't be cancelled within perturbative QCD \cite{Bodwin:1992ye} and one have to account for non perturbative effects of the quarkonia bound state and consider colour octet contributions from the Fock space of the $\chi_c$ meson. As we have already mentioned, these contributions are negligibly small in the region far from the divergence, so in our further analysis we will simply impose a cut-off
\begin{equation}
(p_{\chi} p_g) > \Delta_\chi^2.
\end{equation}
It is interesting to note, that in the considered process there is no any tangible differences between production features of $\chi_{cJ}$ mesons with different $J$. In the following we will not distinguish between different $\chi_{cJ}$ states and only  total feed-down from $J/\psi + \chi_c$ states to the $J/\psi + J/\psi$ production
$$
d\sigma(J/\psi + \chi_c) \equiv \sum_{J=0,1,2} Br(\chi_{cJ}\to J/\psi\gamma) \, \times \, d\sigma(J/\psi + \chi_{cJ})
$$
 will be considered.

A completely different picture is observed for the $gg \to J/\psi + J/\psi + g$ partonic reaction. First of all, contrary to the previous case, all terms with $d_{ABC}$ cancel each other in matrix element, so we can assign positive charge parity to it. Moreover, all diagrams with FSR of type Fig.~\ref{fig_diagrams}b in total are zero and all other diagrams with FSR (Fig.~\ref{fig_diagrams}c) are infrared and collinear stable, which is quite expected for $S$-wave state. On the other hand, ISR diagrams  (Fig.~\ref{fig_diagrams}d) are divergent in the collinear limit. The nature of this divergence is, however,  completely different from the case of $J/\psi + \chi_c$  production: now it can be absorbed into well known DGLAP equations. In our further estimations we consider the process in the kinematical region safe from this divergence:
\begin{equation}
p_T^{J/\psi J/\psi} > \Delta_\psi
\end{equation}
Finally, one should not forget that in the case of $J/\psi + J/\psi$ leading order process (Fig.~\ref{fig_diagrams}a) also gives a contribution.

\begin{figure}
\includegraphics[width=0.5\textwidth]{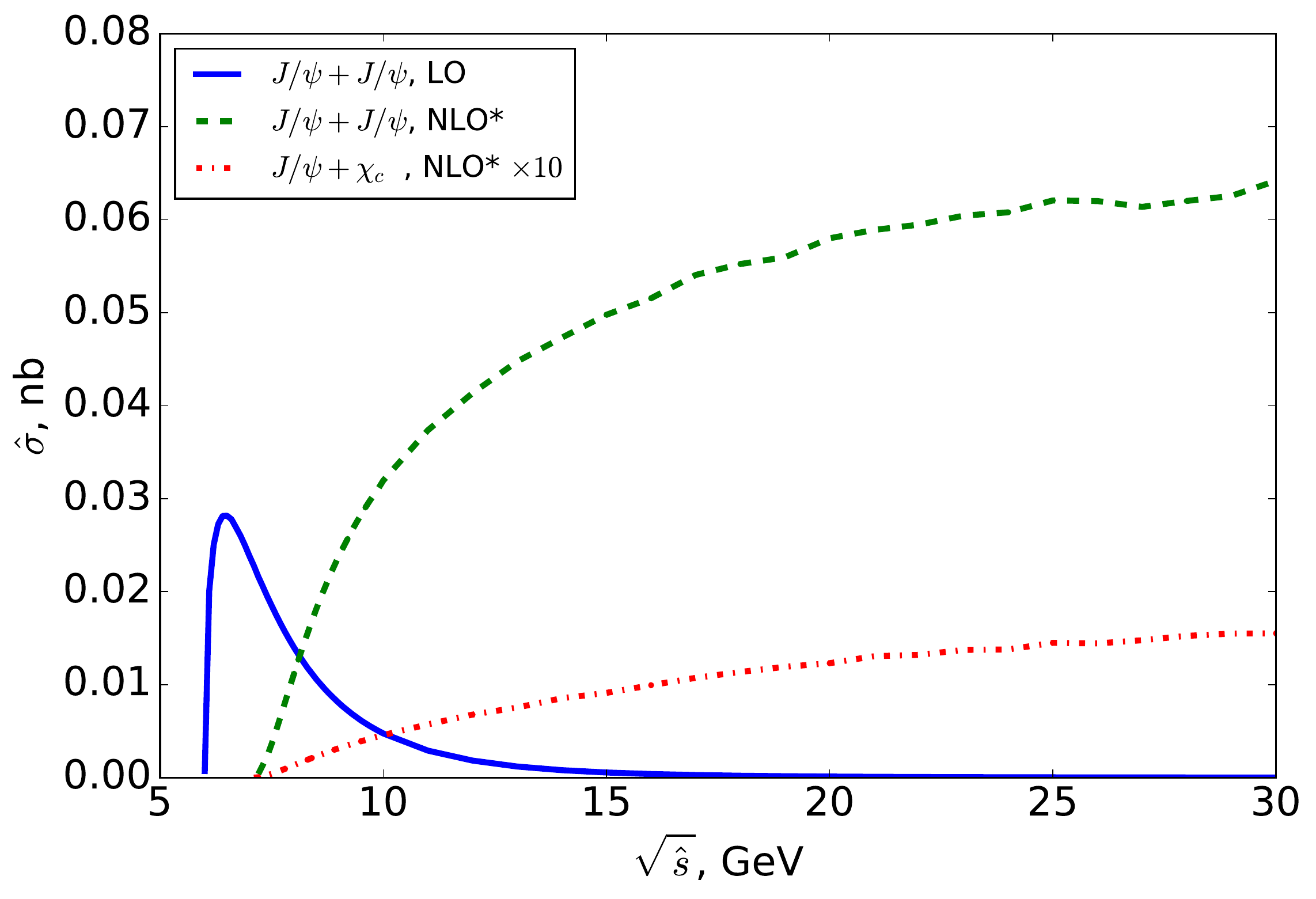}
\caption{Energy dependence of partonic cross section of $gg\to J/\psi + J/\psi$ (solid), $gg\to J/\psi + J/\psi + g$ (dashed), and $gg\to J/\psi + \chi_c + g$ (dot-dashed). The cross sections are evaluated with in the kinematical region $\Delta_\chi = \Delta_\psi = 1\, \GeV$. The cross section of $J/\psi + \chi_c$ is very small, so in the figure it is multiplied by a factor of $10$.
\label{fig_parton_level_ecm}}
\end{figure}

In the following work we will use two sets of cut-off parameters: $\Delta_\chi = \Delta_\psi \equiv \Delta = 1\, \GeV$ and $\Delta_\chi = \Delta_\psi \equiv \Delta = 2m_c  = 3\, \GeV$.
Fig.~\ref{fig_parton_level_ecm} shows the dependence of parton cross section on the total energy of the parton reaction. As one can see, the shapes for the NLO* $J/\psi  + J/\psi$ and $J/\psi  + \chi_c$ are the same, while LO $J/\psi  + J/\psi$ is completely different. The reason is that in LO Feynman diagrams (Fig.~\ref{fig_diagrams}a) momentum of a virtual gluon is fixed and equal to $\sqrt{\hat s}/2$ which leads to the overall cross section decrease as $1/\hat s^2$.

\section{Hadron level}
In the collinear approximation cross section of the hadronic reaction is written in the form:
\begin{eqnarray}
d\sigma &=& \int dx_1 dx_2 f(x_1) f(x_2) d\hat\sigma,
\end{eqnarray}
where $\hat\sigma$ is partonic cross section, $x_{1,2}$ are gluon momentum fractions, and $f_{1,2}(x)$ are PDFs. In our work we use the same scale equal to $\mu^2 = 16\, m_c^2 + (p_T^\psi)^2 + (p_T^\chi)^2$ both for PDFs and  $\alpha_S(\mu)$ . In order to estimate the error due to the choice of the scale, we additionally performed calculations at $\mu/2$ and $2 \mu$. For the PDFs and $\alpha_S$ CT10 parametrization \cite{Lai:2010vv} was used (we have additionally checked CT14 PDFs \cite{Dulat:2015mca} and found that the difference in the results is negligible).

Below we shall present our results at $\sqrt{s}=13 \TeV$ hadronic energy for the LHCb acceptance, i.e. with the rapidity cut $2 < y_{J/\psi} < 4.5$ for each $J/\psi$ meson. Results for other energies and kinematical regions are available at request. Our theoretical predictions for total cross sections are presented in Table~\ref{tab_cross_sections}. For completeness we also present estimates obtained using the DPS model. The total DPS cross section of prompt $J/\psi + J/\psi$ can be obtained via simple formula:
\begin{align}
   \sigma_{\mbox{\scriptsize DPS}} = \frac{1}{2} \frac{\sigma_{J/\psi}^2}{\sigma_{\mbox{\scriptsize eff}}}
   \label{eq:DPS}.
\end{align}
For the single $J/\psi$ cross section and at energy $\sqrt{s}=13\, \TeV$ under LHCb conditions and $\sigma_{\mbox{\scriptsize eff}}$ we take results of \cite{Aaij:2015rla} and \cite{Aaij:2015wpa}:
\begin{align}
	\sigma_{\mbox{\scriptsize eff}} = 18.0\,\mathrm{mb},\qquad
	\sigma_{J/\psi} = 15.3\,\mathrm{\mu b}.
\end{align}


\begin{table}
	\begin{tabular}{|c|c|c|c|c|c|c|c|}
		\hline
		\multirow{2}{*}{Cut} & \multirow{2}{*}{$J/\psi + J/\psi$, LO} & 
		\multirow{2}{*}{$J/\psi + J/\psi$, NLO*} & \multicolumn{4}{|c|}{$J/\psi + \chi_{cJ}$} & 
		\multirow{2}{*}{DPS}  \\
		\cline{4-7}
		&                &                  &  $J = 0$ & $J = 1$ & $J = 2$ & f.-d. & \\
		\hline\hline
		$\Delta=1$ GeV   &   \multirow{2}{*}{$1.29 \,\pm\, 0.02 \, \nb $}   &   $4.47 \,\pm\, 0.79 \, \nb $   &   $11.4 \,\pm\, 1.62 \, \pb $ & $15.4 \,\pm\, 2.26 \, \pb $ & $28.2 \,\pm\, 3.94 \, \pb $ & $10.8 \,\pm\, 1.54 \, \pb$ &$6.2 \, \pm \, 2.1 \, \nb$\\
		$\Delta=3$ GeV   &                                                                       &   $1.68 \,\pm\, 0.33 \, \nb $   &   $2.06 \,\pm\, 0.34 \, \pb $ & $4.19 \,\pm\, 0.71 \, \pb $ & $5.03 \,\pm\, 0.83 \, \pb $ & $2.41 \,\pm\, 0.4 \, \pb$ &$4.0 \, \pm \, 0.6 \, \nb$\\
		\hline
	\end{tabular}
	\caption{Hadronic cross sections in the LHCb acceptance at $\sqrt{s} = 13 \,\TeV$. The last column in $J/\psi + \chi_{cJ}$ section is a total feed-down to $J/\psi + J/\psi$. }
	\label{tab_cross_sections}
\end{table}

\begin{figure}[h]
\includegraphics[width=0.9\textwidth]{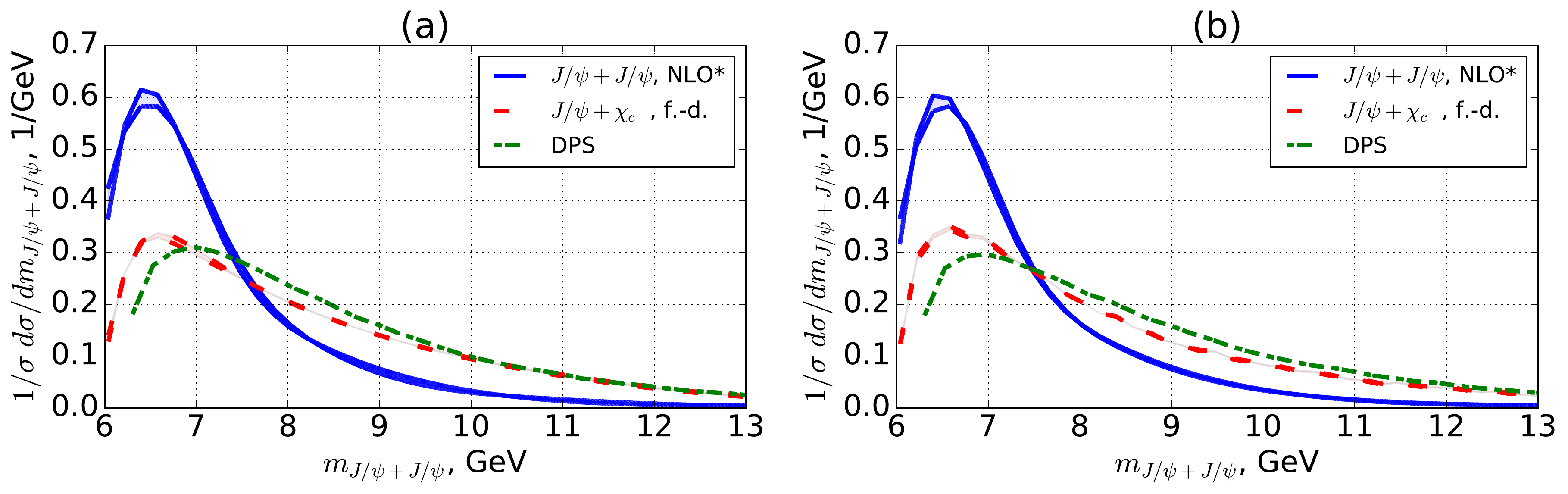}
\caption{Distribution over invariant mass of $J/\psi + J/\psi$ pair for $\Delta=1\, \GeV$ (left) and $\Delta=3\, \GeV$ (right). Solid (blue), dashed (red) and dot-dashed (green) lines co correspond to direct NLO* $J/\psi + J/\psi$, feed-down from $J/\psi + \chi_c$ and DPS mechanisms respectively.}
\label{fig_pp_mPair}
\end{figure}

\begin{figure}[h]
\includegraphics[width=0.9\textwidth]{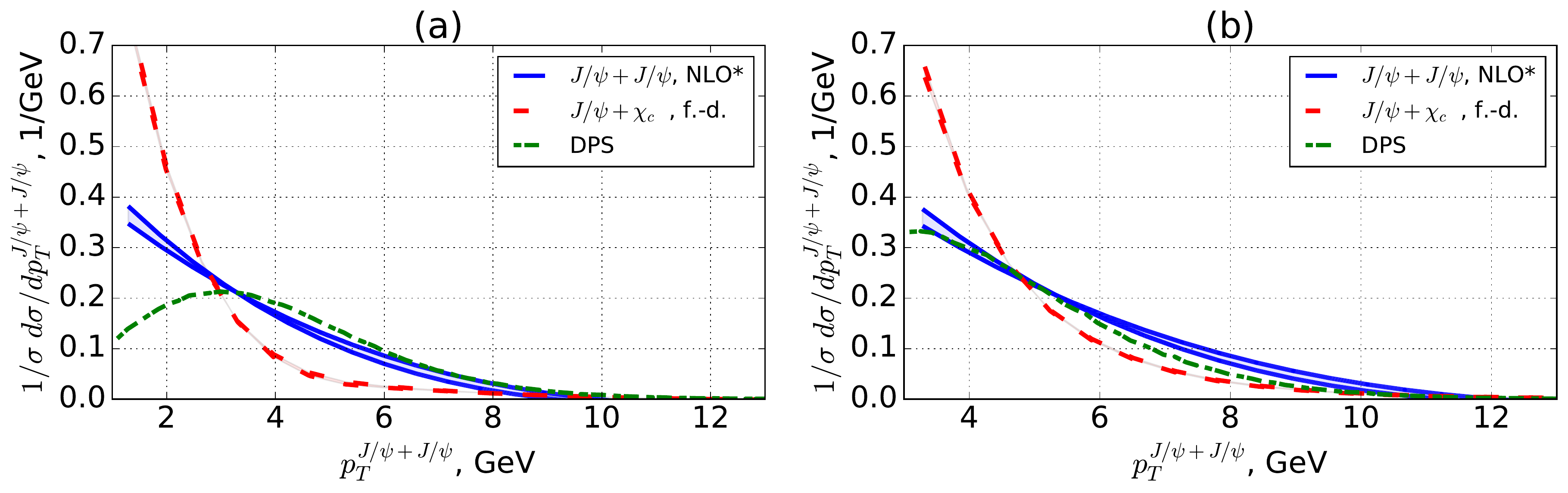}
\caption{Distribution over transverse momenta of a whole $J/\psi + J/\psi$ pair. Notation is the same as in Fig.~\ref{fig_pp_mPair}}
\label{fig_pp_pTpair}
\end{figure}

\begin{figure}
\includegraphics[width=0.9\textwidth]{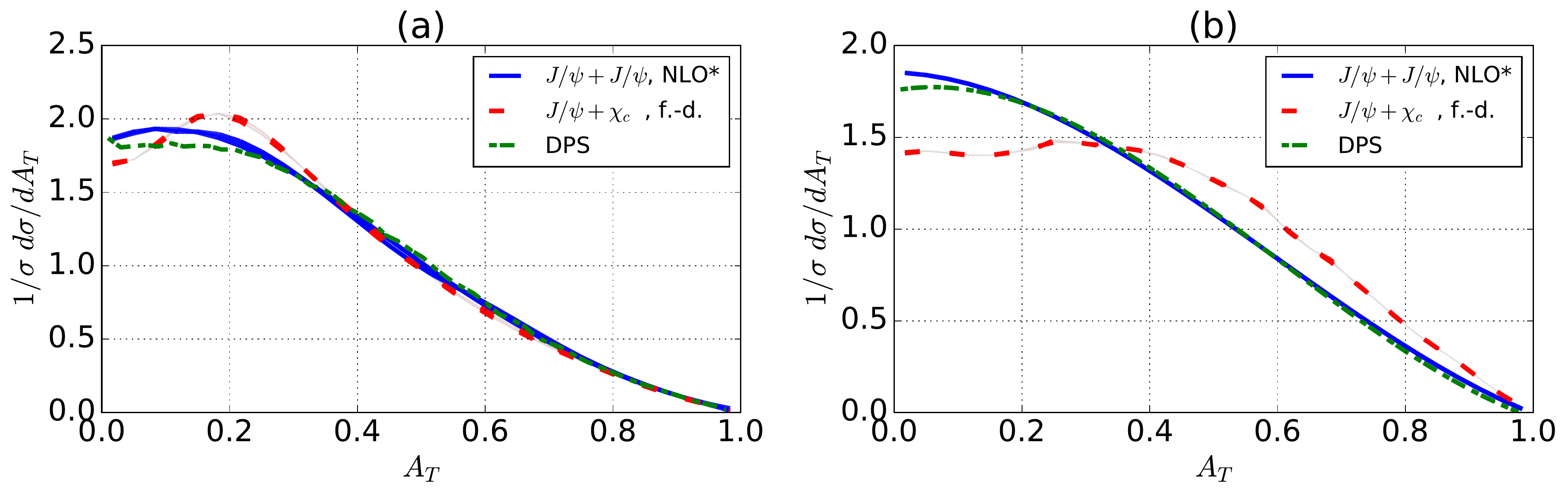}
\caption{Distribution over transverse momentum asymmetry $\mathcal A_T=\left|p_{T1}-p_{T2}\right|/\left|p_{T1}+p_{T2}\right|$. Notation is the same as in Fig.~\ref{fig_pp_mPair}}
\label{pp_delta_pT}
\end{figure}

Now let's turn to the differential distributions. Fig.~\ref{fig_pp_mPair} and \ref{fig_pp_pTpair} show normalized distributions over invariant mass and transverse momenta of $J/\psi$ pair. Form these figures it is clear that the shapes of the distributions are pretty close to each other. It should also be noted that, although the value of the cut-off parameter $\Delta$ affects significantly the values of the cross sections (as it is seen from the Table \ref{tab_cross_sections}), the form of the distributions changes only slightly. The same situation we have for transverse momentum asymmetry $\mathcal A_T=\left|p_{T1}-p_{T2}\right|/\left|p_{T1}+p_{T2}\right|$ shown in Fig.~\ref{pp_delta_pT}.

\begin{figure}[h]
\includegraphics[width=0.9\textwidth]{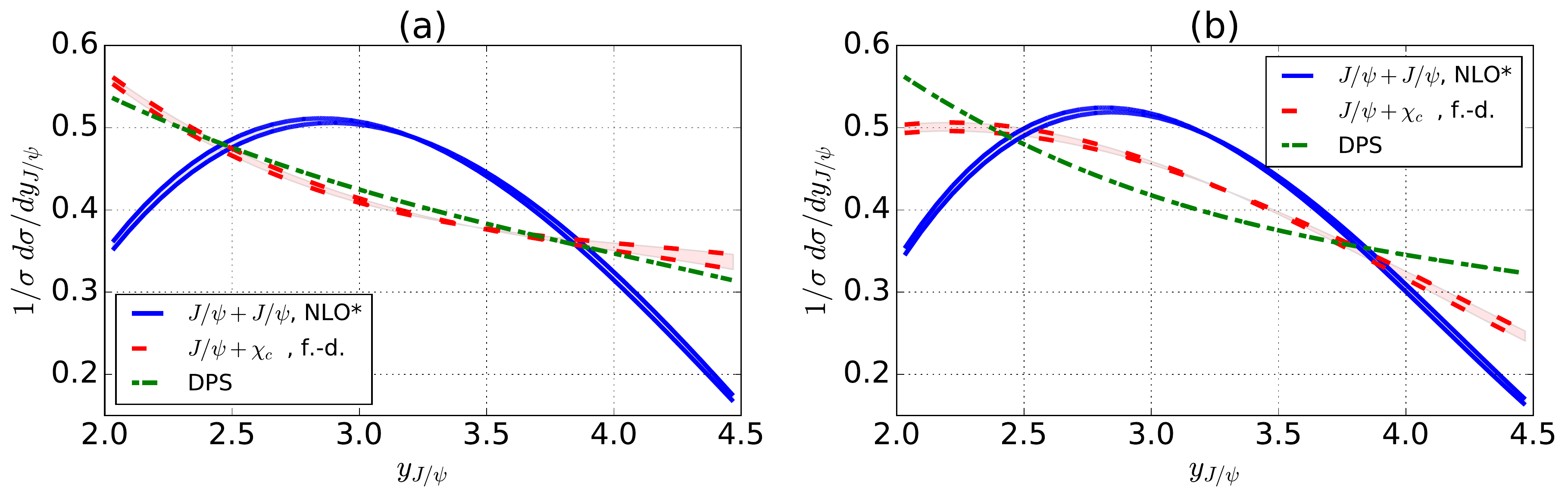}
\caption{Distribution over rapidity of $J/\psi$. Notation is the same as in Fig.~\ref{fig_pp_mPair}}
\label{fig_pp_p1_y}
\end{figure}

\begin{figure}
\includegraphics[width=0.9\textwidth]{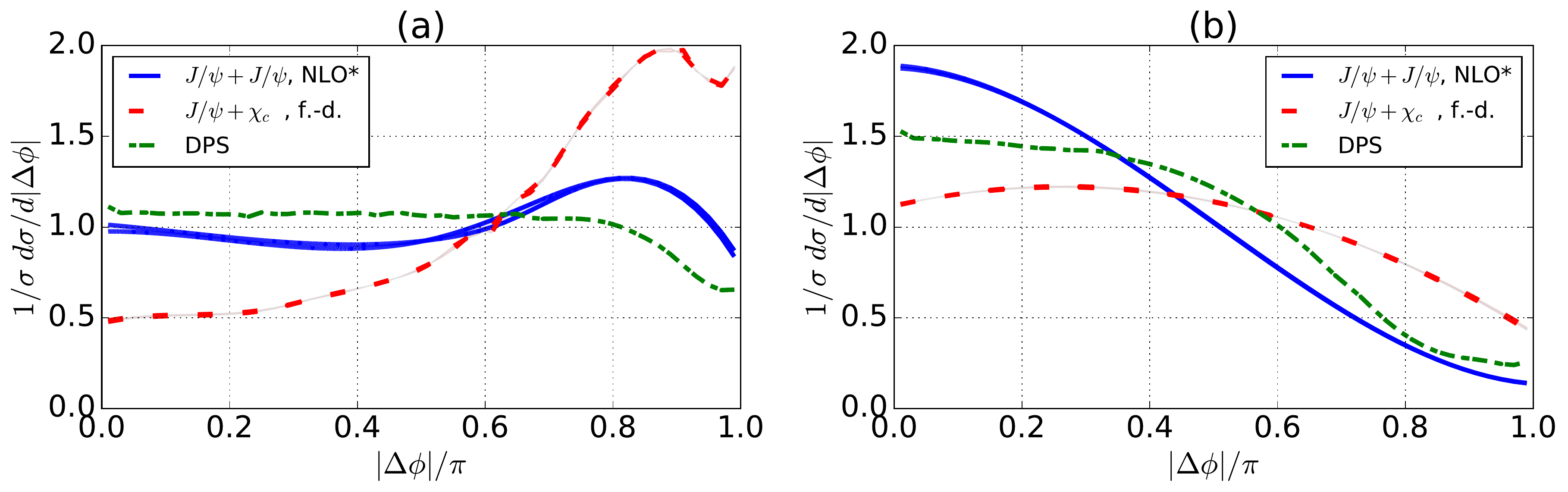}
\caption{Distribution over azimuthal asymmetry $|\Delta\phi|=|\phi_1-\phi_2|$. Notation is the same as in Fig.\ref{fig_pp_mPair}}
\label{fig_pp_delta_phi}
\end{figure}

\begin{figure}
\includegraphics[width=0.9\textwidth]{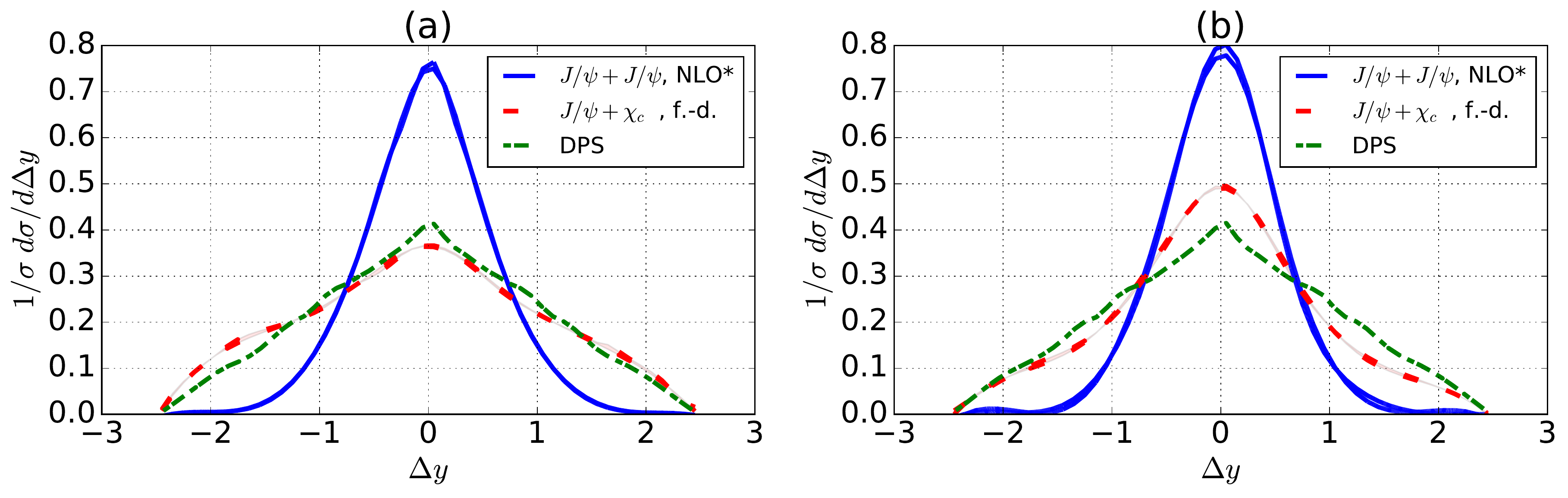}
\caption{Distribution over rapidity asymmetry $\Delta y=y_1-y_2$. Notation is the same as in Fig.~\ref{fig_pp_mPair}}
\label{fig_pp_delta_y}
\end{figure}

Situation changes significantly when we consider distributions over the rapidity of a single $J/\psi$ shown in Fig.~\ref{fig_pp_p1_y}. Here we see that direct NLO* $J/\psi + J/\psi$ production has distinguishing signature. 
Other distinguishing distributions are $|\Delta\phi| = |\phi_1-\phi_2|$ and $\Delta y=y_1-y_2$ shown in Figs \ref{fig_pp_delta_phi} and \ref{fig_pp_delta_y}. It is especially interesting to note how $|\Delta \phi|$ distribution changes in different kinematical regions.

In order to numerically evaluate the difference between form of different distributions one can use, for example, the Pearson correlation coefficient:
\begin{align}
\mathcal P^v_{ij} = 1-\left<\frac{d\sigma_i}{dv} , \frac{d\sigma_j}{dv} \right>,
\end{align}
where $v$ is some kinematical variable and indices $i$ and $j$ mean different channels (prompt, feed-down, DPS). It is clear that if the forms of distributions $d\sigma_i/da$ and $d\sigma_j/da$ coincide, the corresponding correlation $\mathcal P^{v}_{ij}$ is equal to zero, while if they are distinguishable $\mathcal P^{v}_{ij}$ tends to unit. Table~\ref{tab_correlations} summarizes Pearson correlations between different channels at two kinematical regions. As it is seen from the table, there are enough kinematical distributions that can be in principle used to discriminate between contributions from different channels.

\newcolumntype{K}{>{\centering\arraybackslash}p{1cm}}
\begin{table}
\begin{tabular}{|c||l|c|K|K|K|K|K|K|}
\hline
& Combination &  $m_{J/\psi + J/\psi}$  & $p_T^{J/\psi + J/\psi}$  & $|\Delta\phi|$ & $\Delta y$ & $A_T$ & $p_T^{J/\psi}$  & $y_{J/\psi}$\\\cline{2-9}
\multirow{3}{*}{$\Delta = 1\,\GeV$} 
& $J/\psi + J/\psi$ vs $J/\psi + \chi_c$ & 0.10 & 0.02 & \bfseries 0.26 & 0.10 & 0.01 & 0.00 & \bfseries 0.54\\\cline{2-9}
& $J/\psi + J/\psi$ vs DPS & \bfseries 0.21 & \bfseries 0.21 & \bfseries 0.79 & 0.09 & 0.00 & 0.09 & \bfseries 0.55\\\cline{2-9}
& $J/\psi + \chi_c$ vs DPS & 0.03 & \bfseries 0.34 & \bfseries 0.26 & 0.01 & 0.01 & 0.06 & 0.01\\\cline{1-9}

\multirow{3}{*}{$\Delta = 3\,\GeV$}
& $J/\psi + J/\psi$ vs $J/\psi + \chi_c$ & 0.07 & 0.01 & 0.14 & 0.03 & 0.05 & 0.01 & 0.16\\\cline{2-9}
& $J/\psi + J/\psi$ vs DPS & \bfseries 0.20 & 0.02 & 0.04 & 0.11 & 0.00 & 0.05 & 0.15\\\cline{2-9}
& $J/\psi + \chi_c$ vs DPS & 0.04 & 0.06 & 0.04 & 0.03 & 0.04 & 0.02 & 0.01\\\cline{1-9}
\end{tabular}
\caption{Correlation parameters $\mathcal P^{a}_{ij}$. Numbers shown in bold signals that corresponding kinematical distribution can be potentially used to discriminate two corresponding channels. }
\label{tab_correlations}
\end{table}

\section{Conclusion}
In our paper we have studied double charmonium production at LHC with the account of NLO* real gluon emission. The NLO* occurs to be the first non-vanishing contribution to the process of $J/\psi + \chi_c$ production since the LO reaction is forbidden due to the charge parity. On the other hand, in the case of prompt $J/\psi + J/\psi$ production which is allowed at the LO, the NLO* cross section is even larger than the LO. Account for real radiation also naturally solves the problem of strong back-to-back correlation in $2\to2$ kinematics of the LO, thereby providing non trivial distributions over such kinematical variables as total transverse momentum of quarkonia pair, azimuthal and transverse momentum asymmetries.

In Section II it was shown in detail that the production mechanism at the hard scale of partonic reaction is very different for $J/\psi + J/\psi$ and $J/\psi + \chi_c$ processes. In particular by switching between two these final states, we thereby can switch off or switch on different subsets of the underlying Feynman diagrams and test different colour structures of the matrix element. These interesting features can be explained by the charge parity conservation which leads to a non-trivial cancellation between different sets of Feynman diagrams. Additionally, in both processes there are collinear singularities due to the emission of real gluon. However, in the case of $J/\psi + \chi_c$ this singularity arise in final state and its reason in the complicated structure of the Fock space of heavy quarkonia, while in the case of $J/\psi + J/\psi$ singularity arises in the initial state and can be absorbed by a well known Altarelli-Parisi equations. Both singularities in this work were regularized by imposing a cut-off restriction.

In Section III the hadronic process was considered. The absolute value of the $J/\psi + J/\psi$ cross section at NLO* is larger than at LO has the order of several nanobarns. Contrary, the cross section of $J/\psi + \chi_c$ is surprisingly small and has the order of picobarns. For completeness, we also considered production of $J/\psi + J/\psi$ within DPS model. We have considered distributions over different kinematical variables and found that the most promising distributions for discriminating different production channels are the distributions over invariant mass of a pair, transverse momenta of a pair, azimuthal asymmetry, transverse momenta asymmetry and rapidity of a single $J/\psi$. The computer code of the implemented event generator as well as instructions on how it can be run are placed at \href{https://bitbucket.org/PoslavskySV/gggpsichi}{\color{blue}bitbucket.org/PoslavskySV/gggpsichi} and 
\href{https://bitbucket.org/PoslavskySV/gggpsipsi}{\color{blue}bitbucket.org/PoslavskySV/gggpsipsi}.

It could be also interesting to study the effects of real gluon emission for double production of other charmonium states (e.g. $J/\psi\eta_c$, $\chi_{c1}\chi_{c2}$, etc.) and consider various polarization asymmetries in these processes. This is the goal of our future work.

\acknowledgements
The authors would like to thank Dr. Ivan Belayev for fruitful discussions. The work of A.K.L. and A.V.L. was supported by the Russian Foundation of Basic Research grant \#14-02-00096. The work of S.V.P. was supported by the Russian Foundation of Basic Research grant \#16-32-60017.


\end{document}